\documentclass[12pt,a4paper,english]{article}
\usepackage{babel,cite} \usepackage[dvips]{epsfig}

\makeatletter

\usepackage{amssymb,amsthm}

\newcommand{\be}{\begin{equation}} 
\newcommand{\ee}{\end{equation}}
\newcommand{\eq}[1]{(\ref{#1})}

\def\nn{\nonumber} 
\def\bea{\begin{eqnarray}} 
\def\eea{\end{eqnarray}}
\newcommand{\barr}{\begin{array}} 
\newcommand{\earr}{\end{array}}
 
\def\one{\mbox{1 \kern-.59em {\rm l}}}
\def\und#1{\underline{#1}}

\def\({\left(} 
\def\){\right)} 
\def\[{\left[} 
\def\]{\right]}

 \def\d{\delta}

  \def\la{\lambda}

\def\R{{\mathbb R}}  
 \def\one{\mbox{1 \kern-.59em {\rm l}}}

\def\cH{{\cal H}}

\def\bit{\begin{itemize}} \def\eit{\end{itemize}}

\def\diag{\mbox{diag}}

   \def\dd{\partial}

\makeatother
\begin{document}

\renewcommand{\title}[1]{\vspace{10mm}\noindent{\Large{\bf
#1}}\vspace{8mm}} \newcommand{\authors}[1]{\noindent{\large
#1}\vspace{5mm}} \newcommand{\address}[1]{{\itshape #1\vspace{2mm}}}

\begin{flushright}
UWThPh-2006-05
\end{flushright}

\begin{center}

\title{ \Large A nontrivial solvable noncommutative $\phi^3$ model \\[1ex]
in 4 dimensions}

\vskip 3mm

\authors{Harald {\sc Grosse} and 
Harold {\sc Steinacker
}}

\vskip 3mm

\address{Institut f\"ur Theoretische Physik, Universit\"at Wien\\
Boltzmanngasse 5, A-1090 Wien, Austria}

\vskip 1.4cm

\textbf{Abstract}

\vskip 3mm

\begin{minipage}{14cm}%

We study the quantization of the noncommutative selfdual $\phi^3$ model in
4 dimensions, by mapping it to a Kontsevich model. 
The model is shown to be renormalizable, provided one 
additional counterterm is included  
compared to the 2-dimensional case \cite{Grosse:2005ig}
which can be interpreted as divergent shift of the field $\phi$. 
The known results for the Kontsevich model allow to 
obtain the genus expansion of 
the free energy and of any $n$-point function, which is 
finite for each genus after renormalization.
No coupling constant or wavefunction renormalization is required.
A critical coupling is determined, beyond which the model
is unstable. This provides a nontrivial interacting NC
field theory in 4 dimensions.

\end{minipage}

\end{center}

\section{Introduction}

This paper is an extension of our previous work \cite{Grosse:2005ig} 
on the noncommutative (NC) Euclidean selfdual $\phi^3$ model. 
In \cite{Grosse:2005ig} we considered the 2-dimensional case, 
and showed that this model can be renormalized and essentially solved 
using matrix model techniques. This 
was achieved by mapping it to the Kontsevich model.

The map from the selfdual NC $\phi^3$ model to the Kontsevich model 
exists for any even dimension, however the 
eigenvalues and their degeneracy 
of the corresponding Kontsevich model depend on the dimension of the
underlying space.
Therefore the properties of the model and in particular its 
renormalizability must be studied separately in different dimensions.
In the present paper we elaborate the 4-dimensional case.
Generalizing \cite{Grosse:2005ig}, we prove 
renormalizability  and obtain closed
expressions for the genus expansion of the free energy $F$, 
based on results of \cite{Itzykson:1992ya,Kontsevich:1992} 
for the Kontsevich model. 
The general $n$-point functions can then be computed 
by taking derivatives of $F$, in an explicit way
for diagonal entries $\phi_{ii}$
and somewhat implicitly for the general case. 
As an example, we work out  the genus 0 contributions
for the 1-- and 2--point functions. 
 
It turns out that as in the 2-dimensional case, the required renormalization 
is determined by the genus 0 sector only, 
and can be computed explicitly. We show that 
all contributions in a genus expansion of any $n$-point function 
are finite and well-defined 
after renormalization, for finite coupling.  
A linear (tadpole) counterterm must be introduced which 
is linearly divergent, while mass and a further 
counterterm are logarithmically divergent.
The coupling constant does not run as expected. 
This implies but is stronger than perturbative renormalization.
We thus obtain a
fully renormalized and essentially solvable NC field theory
with nontrivial interaction, by applying the rich
structure of the Kontsevich model related to integrable models (KdV flows) and
Virasoro constraints
\cite{Kontsevich:1992,Makeenko:1991ec,Itzykson:1992ya}.
The model is free of UV/IR diseases
due to the confining oscillator potential introduced in 
\cite{Grosse:2003nw,Grosse:2004yu,Grosse:2005da,Langmann:2002cc}.

As in \cite{Grosse:2005ig}, 
these results are obtained starting with purely imaginary coupling 
constants $i\la$, but allow analytic continuation 
to real coupling as long as $|\la|$ is small enough. 
If $\la$ is larger than some critical value, 
the model becomes unstable. We determine the corresponding critical coupling,
which is interpreted as 
instability induced by the finite potential barrier.

This paper is largely parallel to our previous work
\cite{Grosse:2005ig} on the 2-dimensional case. Therefore we will be short
in certain issues which have already been discussed there, and
which apply without change.
Nevertheless, the present paper is essentially self-contained. 
In section \ref{sec:phi3} we define the $\phi^3$ model
under consideration, and rewrite it as 
Kontsevich model. There is one additional counterterm 
compared to \cite{Grosse:2005ig}.
 We then briefly recall the most important facts about the
Kontsevich model in section \ref{sec:kontsevich}.
Renormalization and  finiteness are established in section
\ref{sec:renormaliz}, which is the main result of this paper. 
The  2-point function at genus 0 
is worked out explicitly in section \ref{sec:propagator},
and the computation of the general
$n$-point function is discussed in section \ref{sec:general-correl}.
The critical point is determined in section \ref{sec:critical}. 
 We conclude with a discussion and outlook.

\section{The noncommutative $\phi^3$ model}
\label{sec:phi3}

We consider the 4-dimensional scalar noncommutative 
$\phi^3$ model, with an additional oscillator-type potential 
in order to avoid the problem of UV/IR mixing. 
The model is defined by the action 
\be 
\tilde S = \int_{\R^{4}_{\theta}} \frac 12 \partial_i\phi
\partial_i\phi + \frac {\mu^2}2 \phi^2 + \Omega^2 (\tilde x_i \phi)
(\tilde x_i \phi) + \frac{i\tilde\lambda}{3!}\;\phi^3 
\ee 
on the $4$-dimensional quantum plane, which is generated by self-adjoint
operators\footnote{We ignore possible operator-technical subtleties
here, since the model will be regularized using a cutoff $N$ below.} $x_i$
satisfying the canonical commutation relations
\be 
[x_i,x_j] = i \theta_{ij}, 
\label{CCR}
\ee 
$i,j=1,2,3,4$. We also introduce 
\be 
\tilde x_i = \theta^{-1}_{ij} x_j, \qquad [\tilde x_i,\tilde
x_j] = i \theta^{-1}_{ji}
\ee 
assuming that $\theta_{ij}$ is 
nondegenerate.
The dynamical object is the scalar field
$\phi = \phi^\dagger$, which is  a self-adjoint operator acting
on the representation space $\cH$ of the algebra \eq{CCR}.
The term $\Omega^2 (\tilde x_i \phi)(\tilde x_i \phi) $ is included following 
\cite{Grosse:2003nw,Grosse:2004yu,Grosse:2005da,Langmann:2002cc}, 
making the model covariant under 
Langmann-Szabo duality, and taking care of the UV/IR mixing.
We choose to write the action with an imaginary coupling $i\tilde \la$, 
assuming $\tilde \la$ to be real. The reason is that for real coupling
$\tilde \la' = i\tilde \la$, the potential 
would be unbounded from above and below, 
and the quantization would seem ill-defined. We will see 
however that 
the quantization is completely well-defined for imaginary 
$i\tilde \la$, 
and allows analytic continuation to real $\tilde \la' = i\tilde \la$
in a certain sense which will be made precise below.
Therefore we accept for now that the action $\tilde S$ is not
necessarily  real.

Using the commutation relations \eq{CCR}, the derivatives $\partial_i$
can be written as inner derivatives
$\partial_i f = -i[\tilde x_i,f]$.
Therefore the action can be written as 
\be 
\tilde S = \int -(\tilde
x_i\phi\tilde x_i\phi - \tilde x_i \tilde x_i \phi\phi) + \Omega^2
\tilde x_i \phi \tilde x_i \phi + \frac {\mu^2}2 \phi^2 +
\frac{i\tilde \lambda}{3!}\;\phi^3 
\ee 
using the cyclic property of the
integral. 
For the ``self-dual'' point $\Omega =1$, this action simplifies
further to 
\be
\tilde S = \int (\tilde x_i \tilde x_i + \frac
{\mu^2}2) \phi^2 + \frac{i\tilde \lambda}{3!}\;\phi^3 \,
=\, Tr\Big(\frac 1{2} J \phi^2 +
 \frac{i\lambda}{3!}\;\phi^3 \Big).
\label{action-E} 
\ee
Here we replaced the integral by $\int = (2\pi \theta)^2 Tr$, and
introduce 
\be
J = 2(2\pi \theta)^2 (\sum_i \tilde x_i \tilde x_i + \frac
{\mu^2}2 ),\qquad
\la = (2\pi \theta)^2\tilde \lambda.
\label{const-defs}
\ee 
We  assume that $\theta_{ij}$ has the canonical form 
$\theta_{12} = -\theta_{21} =: \theta \,\,= \theta_{34}= -\theta_{43}$. 
Then $J$ is essentially 
the Hamiltonian of a 2-dimensional quantum mechanical harmonic
oscillator, which in the usual basis of
eigenstates diagonalizes as 
\be 
J|n_1,n_2\rangle 
= 8\pi^2 \theta\, (n_1\;+ n_2\;  + 1
+ \frac {\mu^2\theta}{2})|n_1,n_2\rangle, 
\qquad n_i \in \{0,1,2,...  \}.
\label{J-explicit}
\ee 
To simplify the notation, we will use the convention 
\be
n \equiv (n_1,n_2), \qquad \und{n} \equiv n_1+n_2
\label{n-notation}
\ee
throughout this paper, keeping in mind that $n$ denotes  a double-index.

In order to quantize the theory,  we need to include a linear
counterterm $-Tr (i \la) a\, \phi$ to the action (the explicit factor
$i\la$ is inserted to keep most quantities real), and -- 
as opposed to the 2-dimensional case \cite{Grosse:2005ig} -- 
we must also allow for a divergent
shift 
\be
\phi \to \phi + i\la c
\label{phi-shift}
\ee
of the field $\phi$. 
These counterterms are necessary to ensure that the local 
minimum of the cubic potential remains at the origin after quantization.
The latter shift implies in particular 
that the linear counterterm 
picks up a contribution
$-Tr (i\la)(a+ c J)\phi $ 
from the quadratic term.
Therefore the linear term should be replaced by $-Tr (i\la) A\phi$
where
\be
A = a+c J,
\label{A-def}
\ee
while the other effects of this shift $\phi \to \phi + i \la c$ 
can be absorbed by
a redefinition of the coupling constants 
(which we do not keep track of). 
We are thus led to consider the action
\be 
S = \,Tr \Big( \frac 12 J
\phi^2 + \frac{i \la}{3!}\;\phi^3 - (i\la) A \phi 
- \frac 1{3(i\la)^2}
J^3 -  J A\Big).
\label{action-kontsevich}
\ee 
involving the constants $i\la, \, a,\, c$ and $\mu^2$.
The additional constant terms in \eq{action-kontsevich} 
are introduced for later convenience.
By suitable shifts in the field $\phi$, one can now either eliminate
the linear term or the quadratic term in the action,
\be 
S= Tr \Big( -\frac 1{2 i \la} M^2 \tilde\phi + \frac{i
\la}{3!}\;\tilde\phi^3 \Big) 
= \,Tr \Big(\frac 12 M X^2 + \frac{i\la}{3!}\;X^3 -
\frac 1{3(i\la)^2} M^3 \Big)
\label{action-kontsevich-new}
\ee
where\footnote{for the
  quantization, the integral
for the diagonal elements is then defined via analytical continuation,
and the off-diagonal elements remain hermitian since $J$ is diagonal.}
\be 
\tilde\phi = \phi + \frac 1{i\la} J \, = \, X + \frac 1{i\la} M 
\ee 
and
\bea
M &=& \sqrt{J^2 + 2 (i\la)^2 A} 
= \sqrt{\tilde J^2 + 2 (i\la)^2 a-(i\la)^4 c^2} \label{M-def}\\
\tilde J &=& J + (i\la)^2 c.
\eea 
This has precisely
the form of the Kontsevich model \cite{Kontsevich:1992}. 
Coupling the field linearly to the source will be very useful
for computing correlation functions. 
In particular, this can be compared directly with the 
2-dimensional case considered in \cite{Grosse:2005ig}, 
however $\tilde J_k$ plays now
the role of $J_k$, and a further constant $c$ has been introduced.

\subsection{Quantization and equations of motion}
\label{sec:quantization}

The quantization of the model \eq{action-kontsevich} 
resp. \eq{action-kontsevich-new}
is defined by an integral over
all Hermitian $N^2\times N^2$ matrices $\phi$, where $N$ serves as a UV
cutoff. The partition function is defined as
\be 
Z(M) = \int D\tilde \phi \,\exp(- Tr \Big( -\frac 1{2 i \la} M^2
\tilde\phi 
+ \frac{i\la}{3!}\;\tilde\phi^3 \Big)) = e^{F(M)},
\label{Z-again}
\ee 
which is a function of the eigenvalues of $M$ resp. $\tilde J$.
Since $N$ is finite, we can freely switch between the various 
parametrizations \eq{action-kontsevich}, \eq{action-kontsevich-new}
involving $M$, $J$, $\phi$, or $\tilde\phi$.
Correlators 
or ``$n$-point functions'' are defined through
\be
\langle \phi_{i_1 j_1} ...  \phi_{i_n j_n}\rangle
= \frac 1Z\, \int D \phi \,\exp(- S)\,
\phi_{i_1 j_1} ....  \phi_{i_n j_n},
\label{correl-def}
\ee
keeping in mind that each $i_n$ denotes a double-index \eq{n-notation}.
The (degenerate) 
spectrum of $J$ resp. $M$ is given by \eq{J-explicit} 
resp. \eq{M-def}.
The nontrivial task is to show that all $n$-point functions
have a well-defined 
and hopefully nontrivial limit $N \to \infty$, so that the
``low-energy physics''  is well-defined and independent of the cutoff.

Using the symmetry $Z(M) = Z(U^{-1} M U)$ for $U \in U(N^2)$, 
we can assume that 
$M$ is diagonalized with (ordered) eigenvalues $m_i$. 
In fact, $Z$ depends only on the
eigenvalues of $M^2$ resp. $\tilde J^2$.
There is a residual $U(1)\times U(2) \times U(3) \times ...$ symmetry,
reflecting the degeneracy of $J$. 
This implies certain obvious ``index conservation laws'',
such as $\langle\phi_{kl} \rangle = \delta_{kl} \langle\phi_{ll} \rangle$
etc.

In order to have a well-defined limit $N\to\infty$, 
we must require in particular that the 2-point function 
$\langle\phi_{ij}\phi_{kl}\rangle $ and 
also the one-point function $\langle\phi_{kl} \rangle$ 
have a well-defined limit.
We therefore impose the renormalization conditions 
\bea 
&&\langle\phi_{00} \phi_{00}\rangle = \frac 1{2\pi} \frac 1{\mu_R^2\theta +1}, 
\label{renorm-cond-0}\\ 
&& \langle\phi_{00} \rangle =0 \label{renorm-cond}
\eea 
together with the requirement that the index dependence at least of the
2-point function is nontrivial.
This will uniquely determine the renormalization of $a, c$ and  $\mu^2$, 
while $\la$ 
receives only finite quantum corrections which
will not be computed here.

\paragraph{Quantum equations of motion and correlators.}

Noting that the field $\tilde \phi$ couples linearly to $M^2$ 
resp. $\tilde J^2$ in \eq{Z-again}, one can compute
insertions of a diagonal factor $\tilde
\phi_{kk}$ in a correlator by acting with
the derivative operator 
$2i\la \frac{\dd{}}{\dd \tilde J_k^2}$ 
on $Z$ resp. $F$. More general non-diagonal insertions 
$\tilde \phi_{kl}$ can also be obtained in principle, as discussed in section 
\ref{sec:general-correl}. 
However, using various standard manipulations of the path integral 
\eq{Z-again}
one can derive directly a number of nontrivial identities for the 
$n$-point functions. 
Since their derivation as given in
\cite{Grosse:2005ig} is independent of the eigenvalues of $\tilde J$,
we simply write down the most important identities here, 
noting that $\tilde J$ must be used instead of $J$.
In particular, one finds for the propagator
\be
\langle\tilde \phi_{kl}\tilde \phi_{lk}\rangle = 
 \frac {2i\la}{m_k^2-m_l^2} \langle\tilde\phi_{kk}-\tilde \phi_{ll}  \rangle
\label{Sdyson-3}
\ee
for $k \neq l$ (no sum), where $m_k$ denotes the eigenvalues of $M$. 
Recalling that $\tilde\phi = \phi + \frac 1{i\la}J$, this gives
\bea
\langle\phi_{kl}\phi_{lk}\rangle 
&=& \frac{2}{\tilde J_k + \tilde J_l} 
+ \frac{2i\la}{\tilde J_k^2-\tilde J_l^2} \langle\phi_{kk}-\phi_{ll}  \rangle 
\label{2point-eom}
\eea
noting that $\tilde J_k^2 - \tilde J_l^2 = m_k^2 - m_l^2$ 
using \eq{M-def}.
The first term is the free contribution, and the second the
quantum correction.
This ``only'' requires the 1-point functions
\be
\langle\tilde \phi_{kk}\rangle 
= 2i\la\frac{\dd{}}{\dd m_k^2 } \,\ln \tilde Z(m)  
= \frac 1{i\la} J_k  + \langle\phi_{kk}\rangle
\label{linear-expect}
\ee
which can be obtained from the Kontsevich model, as we will show 
in detail.
Furthermore, one can derive
\be
\frac{m_k^2}{\la^2} = -\langle \tilde \phi_{kk}^2 \rangle 
 -(2i\la) \sum_{l, l\ne k}
{\langle \tilde \phi_{kk}- \tilde \phi_{ll}\rangle \over
  m_k^2-m_l^2}.
\label{master-2}
\ee
These manipulations can be generalized \cite{Grosse:2005ig}.
In particular, certain 3-point functions can be expressed exactly 
in terms of 1- and 2-point functions, etc. We will not repeat these 
considerations here.
Instead, the finiteness of general correlation functions 
will be established directly,
by showing that the appropriate derivatives of the
(connected) generating
function $F(\tilde J) = \ln Z(\tilde J)$ are finite and well-defined 
after renormalization.
This could also be used to demonstrate that the model is not free,
in contrast to \cite{Langmann:2003if}. 
We will not bother to elaborate this, since 
the model will be renormalized for finite coupling, and 
the lowest nontrivial term in an expansion in $\la$ 
is manifestly finite.

\section{Some useful facts for the Kontsevich model}
\label{sec:kontsevich}

The Kontsevich model is defined by
\be
 Z^{Kont}(\tilde M)=e^{F^{Kont}}=
\frac{\int dX \exp\left \{Tr \left(-\frac{\tilde M X^2}{2}
+i \frac{X^3}{6} \right)\right\} }
{\int dX \exp\left\{-Tr \left(\frac{\tilde M X^2}{2}\right)\right\} }
\label{Z-Konts}
\ee
where $\tilde M$ is a given hermitian $N^2 \times N^2$ matrix,
and the integral is over Hermitian $N^2 \times N^2$ matrices $X$.
This model has been introduced by Kontsevich \cite{Kontsevich:1992} 
as a combinatorial way of computing certain topological
quantities (intersection numbers) on moduli spaces of Riemann surfaces
with punctures, which in turn were related to the partition function
of the general one-matrix model by Witten \cite{Witten:1990hr}.
It turns out to have an extremely rich
structure related to integrable models (KdV flows) and
Virasoro constraints,
and was studied using a variety of techniques. For our purpose, the most 
important results are those of 
\cite{Kontsevich:1992,Makeenko:1991ec,Itzykson:1992ya} which provide
explicit expressions for the genus expansion of 
the free energy.
Note that $\la$ can be introduced via
\be
Z^{Kont}(\tilde M) =
\frac{\int dX \exp\left \{Tr \big(-\frac{\tilde M X^2}{2}
+i\frac{X^3}{6} \big)\right\} }
{\int d X \exp\left\{-Tr \big( \frac{ \tilde M X^2}{2}\big)\right\}}
=\frac{\int d\tilde X \exp\left \{Tr \big(-\frac{\ M \tilde X^2}{2}
-i\la\frac{\tilde X^3}{6} \big)\right\} }
{\int d\tilde X \exp\left\{-Tr \big(\frac{M \tilde X^2}{2}\big)\right\}
},
\label{Z-Konts-2}
\ee
where $X = -\la^{1/3}  \tilde X,\, M = \la^{2/3}\tilde  M$, 
which allows to obtain the analytic
continuation in $\la$.

The matrix integral in \eq{Z-Konts} and its large $N$ limit
can be defined rigorously 
in terms of its asymptotic series.
This involves in a crucial way the variables \cite{Kontsevich:1992} 
\be
 t_r := -(2r+1)!!\,\,\theta_{2r+1}, \qquad 
\theta_r := {1\over r} Tr \tilde M^{-r}.
\ee
One can then rigorously define \cite{Kontsevich:1992,Itzykson:1992ya}
the large $N$ limit of the
partition function $Z^{Kont}(\tilde M)$, 
which turns out to be a function of these 
new variables only, $Z^{Kont}(\tilde M) = Z^{Kont}(\theta_i)$. 
More precisely, each order in perturbation theory is a polynomial in
finitely many of the variables $t_r$, and becomes independent of 
$N$ for $N$ large enough. This provides a rigorous
definition of the  $\phi^3$ model. For more details we refer to
\cite{Kontsevich:1992,Itzykson:1992ya,Grosse:2005ig}.

Without renormalization (i.e. for finite or zero $a$), 
$\theta_r$ is linearly divergent for $r=1$, 
logarithmically divergent for $r=2$, 
and finite 
for $r \geq 3$. This is a first indication that the model
requires renormalization.

A further crucial fact is the existence of a  
\paragraph{Genus expansion.}

As usual for matrix models,
one can consider the genus expansion 
\be
\ln Z^{Kont} = F^{Kont} = \sum_{g\geq 0} F^{Kont}_g
\ee
by drawing the Feynman diagrams on a suitable Riemann surface.
In principle, 
this genus expansion can be obtained
as a $\frac 1N$ expansion by introducing
an explicit factor $N$ in the action, 
so that the action takes the form
\be 
S' = \,Tr N ( -\frac 12 M'^2 \phi' +
\frac{1}{3!}\; \phi'^3).
\label{action-kontevich-Nprime}
\ee 
However, it was shown in \cite{Itzykson:1992ya} that
the $F^{Kont}_g$ can also be
computed  using the KdV equations and the Virasoro constraints, 
which allows to find closed expressions
for given $g$. 
It is useful to use the following set of variables:
\be
I_k(u_0,t_i) = \sum_{p \geq 0} t_{k+p} \frac{u_0^p}{p!}
\label{I-k-1}
\ee
where $u_0$ is given by the solution of the implicit equation
\be
u_0 = I_0(u_0,t_i).
\ee
We note that $I_k$ can be resummed as
\be
I_k(u_0,t_i) = - (2k-1)!! \sum_{i} \frac 1{(\tilde m_i^2 - 2
  u_0)^{k+\frac 12}},
\label{I-k-2}
\ee 
in particular
\be
u_0 = -\sum_{i} \frac{1}{\sqrt{\tilde m_i^2-2u_0}} = I_0.
\label{u-constraint}
\ee
These variables turn out to be more useful for our purpose than the 
$t_r$, since the quantities $\tilde m_i^2 - 2 u_0$ will be finite in the
renormalized model, while the $t_r$ are not.
Using the KdV equations, 
\cite{Itzykson:1992ya} find the following explicit formulas:
\bea 
F^{Kont}_0 &=& {u_0^3\over 6}-\sum_{k}{u_0^{k+2}\over k+2}{t_k\over k!}
+\frac 12 \sum_{k}{u_0^{k+1}\over k+1}
    \sum_{a+b=k}{t_a\over a!}{t_b\over b!}  \label{F-0-IZ}\\
F^{Kont}_1 &=& \frac 1{24} \ln \frac 1{1-I_1},  \label{F-0-IZ-1}\\\
F^{Kont}_2 &=& \frac 1{5760}\left[5{I_4\over (1-I_1)^3} +29{I_3 I_2\over
(1-I_1)^4 } +28{I_2^3\over (1-I_1)^5}\right]\ , 
\label{F-higher}
\eea
etc. All $F^{Kont}_g$ with $g \geq 2$ are  given by
{\em finite} sums of polynomials in $I_k/ (1-I_1)^{{2k+1\over 3}}$, the
number of which is $p(3g-3)$ with $p(n)$ being the number of partitions of
$n$.

An alternative form of $F^{Kont}_0$
can be obtained by solving directly the ``master-equation'' 
\eq{master-2} at genus 0. 
This leads to \cite{Makeenko:1991ec}
\bea
F_0^{Kont}&=&\frac{1}{3}\sum_{i} \tilde m_i^3 -
\frac{1}{3}\sum_{i} (\tilde m_i^2-2u_0)^{3/2}
-u_0 \sum_{i}(\tilde m_i^2-2u_0)^{1/2} \nonumber \\
&&+ \frac{u_0^3}{6}-\frac{1}{2}
\sum_{i,k}\ln\left\{\frac{(\tilde m_i^2-2u_0)^{1/2}+(\tilde m_k^2-2u_0)^{1/2}}
{\tilde m_i+\tilde m_k}\right\}
\label{F0Kont}
\eea
which is equivalent to \eq{F-0-IZ} but more useful in our context. 
All sums over double-indices \eq{n-notation}
here and in the following are to be interpreted as 
$$
\sum\limits_i \equiv \sum\limits_{i_1,i_2=0}^{N-1}
$$
truncating the harmonic oscillators in \eq{J-explicit}.
The parameter $u_0$ is given by the implicit equation \eq{u-constraint}.
This constraint can alternatively
be obtained by considering  $F^{Kont}(\tilde m_i;u_0)$
with $u_0$ as an independent variable, since its equation of motion
\be
\frac{\partial}{\partial u_0}F^{Kont}(\tilde m_i;u_0) 
= \frac 12 (u_0 - I_0)^2 =0
\label{constr-implicit}
\ee
reproduces the constraint. After renormalization, 
all sums will be convergent  for the physical observables
as $N\to\infty$.

\section{Applying Kontsevich to the $\phi^3$ model}
\label{sec:appplic}

We need 
\be
Z = Z^{Kont}[\tilde M] Z^{free}[\tilde M]\exp(\frac 1{3(i\la)^2} Tr M^3)
\ee
where
\be
Z^{free}[\tilde M]= e^{F_{free}}=
\int dX \exp\left(-Tr \left(\frac{\tilde M X^2}{2}\right)\right) 
= \prod_i \frac 1{\sqrt{\tilde m_i}}\,\prod_{i<j} \frac 2{\tilde m_i+\tilde m_j}
\ee
up to  irrelevant constants, so that
\be
F_{free} = - \frac 12\sum_{i, j} 
\ln(\tilde m_i+\tilde m_j) \quad (+ const).
\label{F-free}
\ee
Therefore
\bea
F_0 &:=& F_0^{Kont} + F_{free} + \frac 1{3(i\la)^2} Tr M^3 \nn\\
&=&  - \frac{1}{3}\sum_{i} \sqrt{\tilde m_i^2-2u_0}^3
-u_0 \sum_{i}\sqrt{\tilde m_i^2-2u_0} \nonumber \\
&&+ \frac{u_0^3}{6}-\frac{1}{2}
\sum_{i,k}\ln(\sqrt{\tilde m_i^2-2u_0}+\sqrt{\tilde m_k^2-2u_0}).
\label{F0}
\eea
In the present case, the eigenvalues $\tilde m_i$ are given by 
\eq{Z-Konts-2}, \eq{M-def}
\be
\tilde m_i = \la^{-2/3}\sqrt{J_i^2 +2 (i\la)^2 A_i},
\ee
and the model will be ill-defined without renormalization 
since $u_0$ is linearly divergent. However,  note that only 
the combinations $\sqrt{\tilde m_i^2 -2u_0}$ enter in \eq{I-k-2} and \eq{F0}, 
which can be rewritten using \eq{M-def} as 
\be
\sqrt{\tilde m_i^2 -2u_0} 
=\la^{-2/3}\sqrt{ \tilde J_k^2 + 2b}
\ee
where
\be
b = (i\la)^2 a- \la^{4/3}\, u_0 -(i\la)^4 c^2/2 .
\label{b-def}
\ee
The point is that both $\tilde J$ and 
$b$ will be finite after renormalization,
rendering the model well-defined.

The model and therefore the free energy $F$ depends
a priori on the parameters
$a,c,\la$, and $J_k$ resp. $\tilde J_k$ which in turn contains $\mu^2$.
In the form \eq{F0}, new parameters $b$ and $u_0$ 
have appeared, which are implicitly determined by 
\eq{b-def} and the constraint \eq{u-constraint}, 
\be
(b - (i\la)^2 a +(i\la)^4 c^2/2) = -\sum_{i} \frac{(i\la)^2}{\sqrt{\tilde J_i^2 + 2b}}\,\, .
\label{constraint-3}
\ee
Eliminating $u_0$ by \eq{b-def},
the genus 0 contribution to the partition function \eq{F0} 
takes the form
\bea
F_0 &=& \ln Z_{g=0} 
=\frac{(i\la)^{-2}}{3}\sum_{i} \sqrt{\tilde J_i^2+2b}^3
+\Big(a-\frac{b}{(i\la)^2}-(i\la)^2c^2/2\Big)\sum_{i}\sqrt{\tilde J_i^2 +2b} \nn \\
&& \!\!\!\!\!\!\!\! + \frac{(i\la)^{2}}{6}\Big(a-\frac{b}{(i\la)^2}-(i\la)^2c^2/2\Big)^3 
-\frac{1}{2}
\sum_{i,k}\ln\left(\la^{-\frac 23}\sqrt{\tilde J_i^2+2b}
                        +\la^{-\frac 23}\sqrt{\tilde J_k^2+2b}\right).\nn\\
\label{F0tilde-2}
\eea
We consider $F = F(\tilde J^2)$ as a function of 
(the eigenvalues of) $\tilde J^2$ 
from now on, with further parameters $a,c,\la$,
while $b$ is implicitly determined by  \eq{constraint-3}.
Since $\tilde J_k$ only enters through the combination 
$\sqrt{\tilde J_i^2 +2b}$, we note that
the eigenvalues can be analytically continued
as long as this square-root is well-defined. 

We can now compute various $n$-point functions by taking partial 
derivatives of $F = \sum_g F_g$ (where $F_g = F_g^{Kont}$ for $g\geq 1$)  
w.r.t. $\tilde J_k^2$, as indicated in section 
\ref{sec:quantization}. For the ``diagonal'' $n$-point functions
$\langle\tilde\phi_{ii} ... \tilde\phi_{kk}\rangle$, 
this amounts to varying the
eigenvalues of $\tilde J_k^2$.
In doing so, we must be careful to keep the parameters
$a,c$ constant since they determine the model, and note that $b$ depends 
implicitly on $\tilde J_k^2$ through the constraint \eq{constraint-3}.

Some of these computations simplify by the following observation:
The constraint \eq{constraint-3} for $b$ arises automatically 
through the e.o.m as in \eq{constr-implicit}, noting that
\be
\frac{\partial}{\partial  \tilde J_i^2}  F_0(\tilde J_i^2)
 = \frac{\partial}{\partial \tilde J_i^2} F_0(\tilde J_i^2;b)
 +  \frac{\partial}{\partial b} F_0(\tilde J_i^2;b)
  \frac{\partial}{\partial \tilde J_i^2} b 
=\frac{\partial}{\partial \tilde J_i^2} F_0(\tilde J_i^2;b) 
\label{Fkont-partial}
\ee
using
\be
\frac{\partial}{\partial b} \, F_0(\tilde J_i^2;b) 
= -\frac 12 \Big((\frac{b}{(i\la)^2}-a+(i\la)^2c^2/2)  
+ \sum_{i}\frac 1{\sqrt{\tilde J_i^2 + 2 b}}\,\Big)^2 =0.
\label{constr-implicit-2}
\ee
Thus for derivatives of order $\leq 2$ w.r.t. $\tilde J_k^2$, 
we can simply ignore $b$ and treat it as independent variable, 
since the omitted terms \eq{constr-implicit-2} vanish anyway once
the constraint is imposed.

\subsection{Renormalization and finiteness}
\label{sec:renormaliz}

\paragraph{The 1-point function}

We can now determine the required renormalization of $a$ and $c$, by
considering the one-point function. 
Using  \eq{F0tilde-2}, \eq{constr-implicit-2} and \eq{constraint-3}, 
the genus zero
contribution is
\bea
\langle \tilde \phi_{kk}\rangle_{g=0} &=&
2i\la\frac{\dd{}}{\dd \tilde J_k^2 } \,F_0(\tilde J^2) \nn\\
&=& \frac 1{i\la} \sqrt{\tilde J_k^2+2b}  
 + \frac{i\la}{\sqrt{\tilde J_k^2+2b}}
\Big((a-\frac{b}{(i\la)^2}-(i\la)^2c^2/2)  \nn\\
&& \qquad - \sum_{j=0}^N \frac{1}{\sqrt{\tilde J_k^2+2b}+\sqrt{\tilde
     J_j^2+2b}}\Big)   
 + 2i\la\,\frac{\dd{ b}}{\dd \tilde J_k^2 }\,\frac{\dd{ F_0}}{\dd b } \nn\\
&=& \frac 1{i\la}\sqrt{\tilde J_k^2+2b}
 +  \sum_{j}\frac{(i\la)}{\sqrt{\tilde J_k^2+2b}\sqrt{\tilde J_j^2+2b}
   + (\tilde J_j^2+2b)}
\label{F-derivative-1}
\eea
and therefore
\bea
\langle \phi_{kk}\rangle_{g=0} 
&=&\langle \tilde \phi_{kk}\rangle_{g=0} -  \frac{J_k}{i\la} \nn\\
&=& \frac{1}{i\la} (\sqrt{\tilde J_k^2+2b} -\tilde J_k) +
(i\la) c 
+  \sum_{j}\frac{(i\la)}{\sqrt{\tilde J_k^2+2b}\sqrt{\tilde J_j^2+2b}
   + (\tilde J_j^2+2b)}, \nn\\
\label{F-derivative-2}
\eea
which must be finite and well-defined in the limit $N \to \infty$.

As opposed to the 2-dimensional case, we now have to face a 
logarithmic divergence in the sum on the rhs. 
We note that $\tilde J_k$ resp. $J_k$ depends only on the
combination
\be
\und{k}:= k_1+k_2
\label{und-k-def}
\ee
(recall that $k,j,...$ etc. are 2-component indices \eq{n-notation}).
In analogy to the usual strategy in renormalization,
we  consider the Taylor-expansion of 
\bea
f(\und k) &:=& \sum_{j}\frac{1}{\sqrt{\tilde J_k^2+2b}\sqrt{\tilde J_j^2+2b}
   + (\tilde J_j^2+2b)} = f(0) + (\und k)\, f'(0) + ...  \nn\\
 &=:& f(0) + f_R(\und k)   
\label{F-expansion}
\eea
in $\und{k}$, where (keeping only the divergent part)
\bea
f(0) &\approx& \sum_{j}\frac{1}{\tilde J_j^2+2b} \, + const
\approx \frac{1}{(8\pi^2\theta)^2} \, \int_0^N dx_1 dx_2\, 
\frac 1{(x_1+x_2 +const)^2 }  \, + const \nn\\
&\approx& \frac{1}{(8\pi^2\theta)^2} \, \ln(N) 
\eea
up to finite corrections. 
We anticipate here that $\tilde J_k$ \eq{tildeJ-renorm}
and $b$ \eq{b-equation} will be finite
after renormalization.
Only $f(0)$ is divergent in \eq{F-expansion}, while 
all derivative terms $f'(0)$ etc. and in particular $f_R(\und k)$
are finite and well-defined as $N \to \infty$.

This leads to a clear candidate for the appropriate 
scaling of the bare constants $a,\, c, \,\mu^2$ and $\la$: 
the log-divergence in \eq{F-derivative-2} should be absorbed by
$c$, and 
$\mu^2$ should be determined such that $\tilde J_k$ is finite.
This allows to have finite a coupling constant $\la$, 
which does not require 
renormalization. Keeping $b$ also finite determines 
$a$ through the constraint \eq{constraint-3},
rendering the one-point function \eq{F-derivative-2} 
finite and well-defined.
Thus we  set
\be
c = - f(0) +  c'
  \approx -\frac{1}{(8\pi^2\theta)^2} \, \ln(N) +  c'
\label{c-renormaliz}
\ee
where $c'$ is a free finite parameter. 
Furthermore using 
\be
\tilde J_k = J_k + (i\la)^2 c = 8\pi^2 \theta  (\und{k}+1) 
+ (4\pi^2 \theta^2  \mu^2 + (i\la)^2 c)
\ee
we choose according to the above discussion 
\bea
\mu^2 &=& -\frac{(i\la)^2}{4\pi^2 \theta^2}\, c \, + \mu_R^2
= \frac{(i\la)^2 f(0) - (i\la)^2 c'}{4\pi^2 \theta^2}\,  + \mu_R^2\nn\\
&\approx& \frac{(i\la)^2}{256 \pi^6 \theta^4} \, \ln(N) \,
\label{mu-renormaliz}
\eea
where $\mu_R^2 > 0$ is finite, and a free parameter of the model. 
This leads to
\be
\tilde J_k = 8\pi^2 \theta  (\und{k}+1 + \frac{\mu_R^2\theta}2),
\label{tildeJ-renorm}
\ee
which is finite and independent of $N$.

The parameter $b$ is determined by the renormalization conditions
$\langle\phi_{00}\rangle = \langle\tilde
\phi_{00}\rangle-\frac{\tilde J_0}{i\la} =0$  \eq{renorm-cond}.  
At genus 0, this amounts to
\be
\sqrt{\tilde J_0^2 +2b}-\tilde J_0  +(i\la)^2 c' = 0  
\label{b-equation}
\ee
using the above definitions. This has indeed a solution
with finite $b$
as long as $|\frac{(i\la)^2c'}{\tilde J_0}|<1$, 
determining $b$ as a function of $c'$. Note 
that $b = O(\la)^2$ is always analytic in $\la$, starting at second order.
One particular solution is 
\be
c'=0\quad\Rightarrow\,\, b=0
\label{bczero}
\ee
where the formulas take a very simple form.

Since $\tilde J$ and $b$ are now finite and independent of $N$, 
we obtain the renormalized one-point function 
\be
\langle \phi_{kk}\rangle_{g=0} 
= \frac 1{i\la} (\sqrt{\tilde J_k^2+2b}-\tilde J_k) + (i\la) c' 
+ (i\la) f_R(\und k),
\label{F-derivative-4}
\ee
which is finite and well-defined as $N \to \infty$.
Note that there is one additional free parameter compared to the 
2-dimensional case, given by $c'$. 
For the simplest case $c'= b=0$, this simplifies as
\bea
\langle \phi_{kk}\rangle_{g=0} &=&  (i\la) f_R(\und k)
= (i\la) \sum_{j} \frac 1{\tilde J_j}\, 
\Big(\frac{1}{\tilde J_k +\tilde J_j} - \frac{1}{\tilde J_0 + \tilde J_j}\Big)
\label{F-derivative-4simple}
\eea
which coincides with the one-loop result \eq{onepoint-oneloop}; 
we will comment on this fact later.
Finally, $a$ is determined by the constraint \eq{constraint-3},
\bea
(\frac{b}{(i\la)^2} -a +(i\la)^2c^2/2) &=& -\sum_{i} \frac 1{\sqrt{\tilde J_i^2 + 2 b}}
=: - g(b) \label{g-def}\\
&\approx& -\frac{1}{8\pi^2 \theta} \int_0^N dx_1 dx_2\, 
\frac 1{(x_1+x_2 +1)} \approx -\frac{1}{8\pi^2 \theta} N \ln N. \nn
\eea
Therefore 
\be
a  =  g(b) + \frac{b}{(i\la)^2}+ (i\la)^2c^2/2
=  g(0) + (i\la)^2c^2/2 + (finite).
\label{a-renormaliz}
\ee
Thus we can trade $a$ for the implicit parameter $b$, and interpret
$b$ as parametrization of $a$ which is 
determined by the renormalization conditions
$\langle\phi_{00}\rangle =0$.

To summarize, the one-point function is renormalized
by requiring the bare parameters $c,\mu$ and $a$ to scale as in
\eq{c-renormaliz}, \eq{mu-renormaliz}, and \eq{a-renormaliz}.
This leaves 3 independent finite parameters $\la,c'$ and $\mu_R$
of the model. A fourth free parameter $b$ could be introduced by
relaxing the condition $\langle\phi_{00}\rangle =0$.

\paragraph{Diagonal $n$-point functions}

By taking higher derivatives 
of $F_0$ w.r.t. $\tilde J^2$, we obtain  
the genus 0 contribution to the {\em connected part of the} $n$-point 
functions for diagonal entries 
$\langle \tilde \phi_{i_1 i_1} ... \tilde \phi_{i_n i_n}\rangle$. 
Note that the 
(infinite) shift $\tilde \phi = \phi + \frac{J}{i\la}
 = \phi + \frac{\tilde J}{i\la} - (i\la) c$
drops out from the connected $n$-point function for $n \geq 2$, 
therefore for $n \geq 2$ these coincide with
$\langle \phi_{i_1 i_1} ... \phi_{i_n i_n}\rangle$
and should thus be finite. 

Start with the genus 0 contribution. 
To compute higher derivatives of $F_0$ w.r.t. $\tilde J^2$, we also
must take into account the implicit dependence of $b$ on $\tilde J^2$.
We recall that \eq{constr-implicit-2} can be written as
\be
\frac{\partial}{\partial b} \, F_0(\tilde J_i^2;b) 
= -\frac 12 \Big(\frac{b}{\la^2}+a +\la^2c^2/2- \la^{-2/3}\, I_0\,\Big)^2 
\label{constr-implicit-4}
\ee
which vanishes through the constraint \eq{constraint-3}. 
Considering this as a function of the independent variables 
$\tilde J_k^2$ and $b$, we will also encounter 
\bea
\frac{\partial^2}{\partial b^2 } \, F_0(\tilde J_i^2;b) 
&=& -\Big(\frac{b}{\la^2}+a +\la^2c^2/2- \la^{-2/3}\, I_0\,\Big)
(1- \la^{-2/3}\, \frac{\partial}{\partial b} I_0(\tilde J^2,b) ), \nn\\
\frac{\partial^2}{\partial \tilde J_k^2 \partial b  } \, F_0(\tilde J_i^2;b) 
&=&  \Big(\frac{b}{\la^2}+a +\la^2c^2/2- \la^{-2/3}\, I_0\,\Big)
(\la^{-2/3}\, \frac{\partial}{\partial\tilde J_k^2 } I_0(\tilde J^2,b) ) \nn
\eea
which still vanishes due to the constraint, while e.g.
\bea
\frac{\partial^3}{\partial \tilde J_l^2 \partial \tilde J_k^2 \partial b } 
\, F_0 
&=& -\Big(\la^{-2/3}\, \frac{\partial}{\partial\tilde J_l^2 }I_0\,\Big)
(\la^{-2/3}\, \frac{\partial}{\partial \tilde J_k^2} I_0 ) \nn\\
&& -\Big(\frac{b}{\la^2}+a +\la^2c^2/2- \la^{-2/3}\, I_0\,\Big)
(\la^{-2/3}\, \frac{\partial^2}{\partial\tilde J_l^2\partial \tilde
  J_k^2} I_0)
\nn
\eea
is non-vanishing, and similarly for higher derivatives. 
We will show below that
all these terms are finite as $N \to \infty$ provided $\la$ is small enough. 

Consider the quantities involved in more detail. 
First, all the
\be
I_p(\tilde J_k^2,b) 
= -(2p-1)!! \la^{2(2p+1)/3}
\sum_{i}\frac 1{(\tilde J_i^2 + 2b)^{p+\frac 12}} 
\label{I-p-sum}
\ee
are finite (i.e. convergent as $N \to \infty$)
for $p \geq 1$. In particular, we note that 
$|I_1| <1$ provided the coupling $\la$ is small enough.
Furthermore, the constraint \eq{constraint-3} implies
\be
\frac{\partial}{\partial \tilde J_k^2} b
=  -\frac {\la^2}{2(\tilde J_k^2 + 2b)^{3/2}} 
- \la^2\Big(\sum_{i} \frac {1}{(\tilde J_i^2 + 2b)^{3/2}} \Big)
\frac{\partial}{\partial \tilde J_k^2} b,
\ee
hence
\be
\frac{\partial}{\partial \tilde J_k^2} b
= -\frac {\la^2}{2(\tilde J_k^2 + 2b)^{3/2}}\, \frac {1}{1 - I_1}
\ee
which is finite for small $|\la|$
since we can assume $|I_1|<1$ as just shown.
Finally, we need
\bea
\frac{\partial}{\partial b} I_p(\tilde J,b) &=& - \la^{-4/3}\,
I_{p+1}, \nn\\
\frac{\partial}{\partial \tilde J_k^2}  I_p(\tilde J,b) &=&  
\frac{(2p+1)!! \la^{2(2p+1)/3}}{2(\tilde J_k^2 + 2b)^{p+\frac 12}}.
\eea
which are all finite as $N \to \infty$ provided $\la$ is small enough, 
for any $p \geq 0$. The same is obviously true for all higher
derivatives. 
Furthermore, we see explicitly 
that any higher derivative of the r.h.s. in \eq{F-derivative-1}
w.r.t. $\tilde J_k^2$ or $b$ 
is also manifestly finite. Combining all this, we have shown that all 
relevant
derivatives of $F_0$ are finite and have a well-defined limit $N \to
\infty$, provided $\la$ is small enough, the constraint \eq{constraint-3} is
imposed,
and the parameters $a,c, \mu^2$ are renormalized as in 
\eq{c-renormaliz}, \eq{mu-renormaliz},
and \eq{a-renormaliz}.
This establishes that the genus 0 
contribution to the 
correlation functions for diagonal entries 
$\langle \phi_{kk} ... \phi_{ll}\rangle$ is finite and well-defined.

\paragraph{Higher genus contributions.}

It is easy to see that the higher genus contributions are also finite
under the same conditions. 
Indeed, in view of the  structure of the higher genus contributions $F_g$ 
stated below \eq{F-higher} as found by \cite{Itzykson:1992ya}, this
follows immediately form the above considerations.
Thus we have established 
\begin{itemize}
\item[] {\bf Theorem 1} {\em  All derivatives of $F_g$  w.r.t. 
$\tilde J_k^2$ for $g \geq 0$ as well as all $F_g$ for $g \geq 1$
are finite and have a well-defined limit $N \to \infty$, 
provided $\la$ is small enough, the constraint \eq{constraint-3} is
imposed,
and the parameters $a,c, \mu^2$ are renormalized as in 
\eq{c-renormaliz}, \eq{mu-renormaliz},
and \eq{a-renormaliz}.}
\end{itemize}
The precise condition for $\la$ is that $|I_1|<1$. The
limiting case $I_1=1$ will be identified below as an instability of
the model.

Since the connected $n$-point functions are given by the derivatives
of $F = \sum_{g \geq 0} F_g$ w.r.t. $\tilde J_k^2$, this implies that
all contributions in a genus expansion of the
correlation functions for diagonal entries 
$\langle \phi_{kk} ... \phi_{ll}\rangle$ are finite and well-defined. 
The general non-diagonal correlation functions are
discussed in section \ref{sec:general-correl}, 
and also turn out to be finite 
for arbitrary genus $g$ under the same conditions.
Putting these results together we have
established renormalizability of the model
to all orders in a genus expansion, i.e.
\begin{itemize}
\item[]{\bf Theorem 2} 
{\em The (connected) genus $g$ contribution 
to any given $n$-point function is finite and has a 
well-defined limit $N \to \infty$ for all $g$, under the above conditions.}
\end{itemize}   
Moreover, they can in principle 
be computed explicitly using the above formulas.
In particular, since any contribution to 
$F_g$  has order at least $\la^{4g-2}$, 
this implies renormalizability of 
the perturbative expansion to any order in $\la$. This is 
certainly expected in view of the results in
\cite{Grosse:2003nw,Grosse:2004yu,Grosse:2005da}.

Note in particular that we did not have to specify whether $\la$ is 
real, or $i\la$ is real, etc. Rather, all genus $g$ contributions 
are analytic in $\la$ provided $|\la|$ is small enough such that 
$|I_1|<1$ holds. Therefore we have obtained a definition of the 
NC $\phi^3$ model also for real coupling $\la$, under the
above conditions.

It is also worth pointing out that only the genus 0 contribution 
requires renormalization, while all higher genus contributions are 
finite. This is very interesting because the genus 0 contribution
can be obtained by various techniques in more general models, 
see also \cite{Steinacker:2005wj,Steinacker:2005tf}. This is due 
to the presence of the oscillator-like potential in the action,
which suppress the UV/IR mixing originating from higher genus diagrams.

The parameters $\mu_R^2, \la$ and $c'$ are the free moduli of the 
model, which can be interpreted as mass, coupling, and a further
parameter which was introduced by renormalization.

\subsection{The 2-point function at genus 0}
\label{sec:propagator}

\subsubsection{$\langle\phi_{kl}\phi_{lk}\rangle$.}

We can use \eq{2point-eom} to obtain the genus 0 contribution 
to the 2-point function
$\langle\phi_{kl}\phi_{lk}\rangle$ for $k \neq l$. 
Using \eq{F-derivative-4} we obtain
\bea
\langle \phi_{kl}\phi_{lk}\rangle_{g=0} 
&=& 2\, \frac{\sqrt{\tilde J_k^2+2b} - \sqrt{\tilde J_l^2+2b} 
+ (i\la)^2\,(f_R(\und k) - f_R(\und l))}
{\tilde J_k^2-\tilde J_l^2} .
\label{2point-explicit}
\eea
Note that $\la$ enters also implicitly through $b$ \eq{b-equation}.
For the simplest case $b=0$, this simplifies to
\bea
\langle \phi_{kl}\phi_{lk}\rangle_{g=0} 
&=& \frac 2{\tilde J_k+\tilde J_l} + 
2(i\la)^2\, \frac{f_R(\und k) - f_R(\und l)}
{\tilde J_k^2-\tilde J_l^2}  \nn\\
&=& \frac 2{\tilde J_k+\tilde J_l} 
 - (i\la)^2\,\frac 2{\tilde J_k+\tilde J_l}\sum_{j} 
\frac 1{\tilde J_j(\tilde J_k + \tilde J_j)(\tilde J_l + \tilde J_j)}.  
\label{2point-explicit-2}
\eea
The first term is the free propagator,
while the second term has somewhat stronger decay properties.
This can be rewritten in a more suggestive way using the identity
\be
\frac{\langle \phi_{kk} - \phi_{ll}\rangle}{\tilde J_k^2 - \tilde J_l^2}
+ \frac{\langle \phi_{kk} + \phi_{ll}\rangle}{(\tilde J_k + \tilde J_l)^2}
= 2\,\frac{\tilde J_k\langle \phi_{kk}\rangle - \tilde J_l\langle \phi_{ll}\rangle}
{(\tilde J_k + \tilde J_l)^2(\tilde J_k-\tilde J_l)},
\ee
which for $b=0$ and using \eq{F-derivative-4simple},
\eq{c-renormaliz} leads to
\be
\langle \phi_{kl}\phi_{lk}\rangle_{g=0} 
=\frac 2{\tilde J_k+\tilde J_l} 
- 2(i\la)\,\frac{\langle \phi_{kk}+\phi_{ll}\rangle}{(\tilde J_k+\tilde J_l)^2}
+ \frac{4(i\la)^2}{(\tilde J_k + \tilde J_l)^2}\,
\Big(\sum_j \frac{1}{(\tilde J_k + \tilde J_j)(\tilde J_l + \tilde
  J_j)} +c \Big).
\ee
Comparing with the perturbative computation \eq{2point-eom-2-pert},
this means that 
the one-loop contribution to the propagator gives the exact result
for the genus 0 sector for the case $b=0$, noting that 
$\la^2 c = \d J$ using \eq{mu-renormaliz} and \eq{deltaJ-def}.
This remarkable property can be traced back to \eq{F0tilde-2},
which is also exact at order $\la^2$ for $b=0$. 
For higher genus, this is no longer the case.

\subsubsection{$\langle\phi_{ll}\phi_{kk}\rangle$.}

As a further example, consider the 2-point function 
$\langle\phi_{ll}\phi_{kk}\rangle$ for $k \neq l$, 
which vanishes in the free case.
To compute it from the effective action, we need in principle
\bea
\langle\tilde\phi_{ll}\tilde\phi_{kk}\rangle - \langle\tilde\phi_{kk}\rangle
\langle\tilde\phi_{ll}\rangle 
&=& 2i\la\frac{\partial}{\partial \tilde J_l^2}
2i\la\frac{\partial}{\partial \tilde J_k^2}(F_0 + F_1 + ...) .
\eea
Even though this corresponds to a nonplanar diagram with external legs, 
it is  obtained by
taking derivatives of a closed genus 0 ring diagram.
Therefore we expect that only $F_0$ will contribute, 
and indeed the derivatives of $F_1$ contribute only to order $\la^{4}$.
We need 
\bea
2i\la\frac{\partial}{\partial \tilde J_l^2}
2i\la\frac{\partial}{\partial \tilde J_k^2} F_0 
&=& - \frac{2(i\la)^2}{\sqrt{\tilde J_k^2+2b}}
{\tilde J_l^2}\frac{\partial}{\partial \tilde J_l^2}
 \left(\sum_{j} \frac{1}{\sqrt{\tilde J_k^2+2b}
   +\sqrt{\tilde J_j^2+2b}} \right) \nn\\
&=& (i\la)^2\frac{1}{\sqrt{\tilde J_k^2+2b}}\frac{1}{\sqrt{\tilde J_l^2+2b}}
 \left(\frac{1}{\sqrt{\tilde J_k^2+2b} +\sqrt{\tilde J_l^2+2b}}\right)^2  
\eea
using again \eq{constr-implicit-2}.
Therefore to lowest order we obtain
\bea
\langle\phi_{ll}\phi_{kk}\rangle 
= \langle\phi_{kk}\rangle \langle\phi_{ll}\rangle 
 + \frac{(i\la)^2}{\tilde J_k\, \tilde J_l}\left(\frac{1}{\tilde J_k + \tilde J_l}\right)^2,
\label{phil-phik-kont}
\eea
in complete agreement with the perturbative computation 
\eq{2point-nonplanar}.

\subsection{General $n$-point functions}
\label{sec:general-correl}

Finally we show that all contributions in the genus expansion (and therefore
perturbative expansion) of the expectation values of any $n$-point
functions of the form
\be
\langle \phi_{i_1 j_1} ....  \phi_{i_n j_n}\rangle 
\label{correl}
\ee
have a well-defined and finite limit as $N \to \infty$
provided $b$ is finite, which means
that the model is fully renormalized. The argument is a generalization
of the one given in \cite{Grosse:2005ig}, taking
into account the degeneracy of $J$.

In view of \eq{Z-again},
the insertion of a factor $\tilde\phi_{ij}$ can be obtained by acting
with the derivative operator 
$2i\la\frac{\partial}{\partial \tilde J^2_{ij}}$ on $Z(\tilde J^2)\,$, resp. 
$F_g(\tilde J^2)$ for fixed genus $g$. 
Now any given correlation function of type \eq{correl}
involves only a finite set of indices $i,j,...$ . Thus taking 
derivatives w.r.t. $\tilde J^2_{ij}$ amounts to considering  
matrices $\tilde J$ of the form 
\be
\tilde J = \left(\begin{array}{ccc} \diag(\tilde J_{1},... \tilde
    J_k) + \delta \tilde J_{k\times k} &\vline & 0 \\
                   \hline 
                  0   & \vline & \diag(\tilde J_{k+1},... \tilde J_N)\end{array}\right),
\label{J-block}
\ee
where $k$ is chosen large enough such that all required variations
are accommodated  by the general hermitian $k\times k$ matrix
\be
\tilde J_{k\times k}:= \left(\diag(\tilde J_{1},... \tilde
  J_k)+\delta \tilde J_{k\times k}\right)
\ee
in \eq{J-block},  while
the higher eigenvalues $\tilde J_{k+1},... \tilde J_N$ are fixed and given by 
\eq{tildeJ-renorm}. Therefore we can restrict ourselves to 
this $k \times k$ matrix, which is independent of $N$.
As was shown in section \ref{sec:renormaliz}, all $F_g$ 
are in the limit $N \to \infty$ 
smooth (in fact analytic) symmetric functions of the first $k$
eigenvalues squared, hence of the eigenvalues 
of $(\tilde J_{k\times k})^2$. Such a function 
can always be written as a smooth (analytic) function 
of some basis of symmetric polynomials in the $\tilde J_a^2$,
in particular
\be
F_g(\tilde J_1^2, ..., \tilde J_k^2) = f_g(Tr(\tilde J^2_{k\times
  k}), ..., Tr(\tilde J^{2k}_{k\times k})).
\label{F-g-symm}
\ee
This can be seen by approximating the analytic function
$F_g(z_1, .., z_k)$ at the point $z_i = \tilde J_i^2$
by a totally symmetric polynomial in the $z_i$,
which correctly reproduces the partial derivatives up to some order $n$. 
According to a well-known theorem, 
that polynomial can be rewritten as polynomial in the 
elementary symmetric polynomials, or equivalently 
as a polynomial in the variables 
$s_n:=\sum z_i^n$, $n=1,2, ..., k$. This amounts to the
rhs of \eq{F-g-symm}.

In the form \eq{F-g-symm}, it is obvious that all partial derivatives
$\frac{\partial}{\partial \tilde J^2_{ij}}$ of $F_g$ exist to any given order,
and could be worked out in principle.
This completes the proof that each genus $g$ contribution to 
the general (connected) 
correlators $\langle \phi_{i_1 j_1} ....  \phi_{i_n j_n}\rangle $
is finite and convergent as $N \to \infty$.
This implies in particular (but is stronger than) renormalizability of 
the perturbative expansion to any order in $\la$.

\subsection{Approximation formulas for finite coupling}
\label{sec:asymptotic}

In this section we derive some closed formulas which 
are appropriate for finite coupling $\la$, in the large 
$N$ limit. They will be needed in particular to derive the 
critical line in section \ref{sec:critical}. For simplicity 
we only consider the case $b = c' = 0$ \eq{bczero}. 

We approximate the various sums by integrals. This gives 
\bea
I_0  &=& - \la^{2/3}
\sum_{i}\frac 1{(\tilde J_i^2)^{\frac 12}} \,\,
\approx -\frac{\la^{2/3}}{8\pi^2 \theta} \int_{x_0}^{x_N} dx 
\int_{y_0}^{y_N} dy\, 
\frac 1{(x+y +1)} \nn\\
&\approx& -\frac{ \la^{2/3}}{8\pi^2 \theta}\, N \ln N,
\eea
where 
\be
x_n =  (n+ \frac{1+\mu_R^2\theta}2), \quad
dx = dn \, .
\label{xn-def}
\ee
Similarly,  
\bea
I_1 &=&  - \la^{2}
\sum_i\frac 1{(\tilde J_i^2)^{\frac 32}} 
\approx -\frac{\la^{2}}{(8\pi^2 \theta)^3} \int_{x_0}^{x_N} dx 
\int_{y_0}^{y_N} dy\, 
\frac 1{(x+y +1)^3} \nn\\
&\approx&  -\frac{\la^{2}}{2(8\pi^2 \theta)^3} \int_{x_0}^{x_N} dx
\frac 1{(x+ (\frac{1+\mu_R^2\theta}2) +1)^2} 
\approx  -\frac{\la^{2}}{2(8\pi^2 \theta)^3}
\frac 1{\mu_R^2\theta +2}.
\label{I1-asymptot}
\eea
Similarly, all $I_p$ can be approximated by 
elementary, convergent integrals.

\subsection{Critical line and instability.}
\label{sec:critical}

We have seen that for small enough coupling $|\la|$, 
the free energy $F = F_0 + F_1 + ...$
is regular and  finite for any given genus in the renormalized
model, since all $I_k$ with $k \geq 1$
are finite {\em provided} $I_1 \neq 1$. 

However, as is manifest in the explicit formulas for $F_g$ at higher
genus \eq{F-0-IZ-1} ff., there is a singularity at $I_1=1$.
Using \eq{I1-asymptot}, 
this critical point is given for $c' = b = 0$ by
\be
\mu_R^2\theta +2 = \frac{(i\la)^{2}}{2(8\pi^2 \theta)^3}.
\label{critical-1}
\ee 
This\footnote{Recall that this is obtained
imposing the renormalization conditions $\langle \phi_{00}\rangle =0$ 
at genus 0. Note also that the rhs is indeed dimensionless.}
is similar to the 2-dimensional case \cite{Grosse:2005ig}, and
 indicates that for the $\phi^3$ model with real coupling 
constant $\la' = i\la$ stronger than this critical coupling, 
the model becomes unstable.
This is very reasonable, since the potential is unbounded, and the 
potential barrier around the local minimum becomes weaker 
for stronger coupling. Therefore this critical line
could be interpreted as 
the point where the quantum fluctuations of $\phi$ 
become large enough to see the global instability, so that the 
field ``spills over'' the potential barrier. 
Similar transitions 
for a cubic potential are known e.g. for
the ordinary matrix models, but may also be relevant 
in the context of string field theory
and tachyon condensation \cite{Sen:1999nx,Gaiotto:2003yb}.
In particular, it is interesting to note that this singularity
occurs simultaneously for each given genus, which suggest
that some double-scaling limit near this 
critical point can be taken, again in analogy with the 
usual matrix models (for a review, see
e.g. \cite{DiFrancesco:1993nw}). Again, such a 
scaling limit for the 
Kontsevich model is discussed in \cite{Itzykson:1992ya}. We leave this 
issue for future work.

\section{Perturbative computations}
\label{sec:perturbative}

We write the action \eq{action-kontsevich} as
\bea
\tilde S &=& Tr \Big(\frac 14 (J \phi^2 + \phi^2 J)
+ \frac{i\la}{3!}\;\phi^3 - (i\la) A \phi \Big) \nn\\
&=&  Tr( \frac 12 \phi^i_j \; (G_R)^{j;l}_{i;k}\;  \phi^k_l 
+ \frac{i\la}{3!}\;\phi^3 - (i\la) A \phi
+ \frac 14 (\d J \phi^2 + \phi^2 \d J))
\eea
where the finite (renormalized) kinetic term 
$ (G_R)^{j;l}_{i;k}  = \frac 12 \delta^i_l \delta^k_j (J^R_i+J^R_j)$ 
defines the propagator
\be
\Delta^{i; k}_{j;l} = \langle \phi^i_j \phi^k_l\rangle 
 = \delta^i_l \delta^k_j \frac 2{J^R_i+J^R_j}
= \delta^i_l\delta^k_j\frac {1/(4\pi^2 \theta)}{\und{i} +\und{j} + (\mu_R^2\theta+2)},
\label{propagator}
\ee
and we use again the notation \eq{und-k-def}
$\und{n} = n_1+n_2$.
This
corresponds to the finite (renormalized) matrix
\be
J^R |n_1,n_2\rangle 
= 8\pi^2 \theta(\und{n}+1+\frac{\mu_R^2\theta}2) |n_1,n_2\rangle.
\ee
The remaining 
\be
\d J |n_1,n_2\rangle 
= (8\pi^2 \theta \frac{\d\mu^2\theta}2) |n_1,n_2\rangle
\label{deltaJ-def}
\ee
is part of the counter-term, where
$\d\mu^2 = (\mu^2 - \mu_R^2)$.

\paragraph{1-point function}
The one-loop contribution to the 1-point function gives
\bea
\langle \phi_{ii} \rangle 
&=&  \frac{i\la}{J^R_i} A_i - \frac{i\la}2\,  
\frac 1{J^R_i} \sum_{k}  \frac 2{J^R_i+J^R_k} \quad  + O(\la^2) \nn\\
&=& -\frac{i\la}{J^R_i}\left(- A_i 
+ \frac 1{8\pi^2\theta} \sum_{k} 
 \frac 1{\und{i} +\und{k} + 2+\mu_R^2\theta}\right)\quad  + O(\la^2).
\label{onepoint-oneloop}
\eea
To proceed, we expand
\be
h(\und i) := \frac 1{8\pi^2\theta}\sum_{k} \frac 1{\und{i} +\und{k} +2+ \mu_R^2\theta}
 = h(0) + (\und{i})\, h'(0) + h_R(\und i)
\ee
where
\bea
h(0) &=& \frac 1{8\pi^2\theta}\sum_{k} \frac 1{\und{k} + \mu_R^2\theta+2} 
\sim \frac 1{8\pi^2\theta}\, N\log N, \nn\\
h'(0) &=& - \frac 1{8\pi^2\theta}\sum_{k} \frac 1{(\und{k} +
  \mu_R^2\theta+2)^2} 
\sim - \frac 1{8\pi^2\theta}\, \log N,
\eea
and $h_R(\und i)$ 
is a finite nontrivial function of $\und{i}$.

We note in particular that the $i$- dependent term 
$(\und{i})\, h'(0)$ in 
\eq{onepoint-oneloop} forces us to introduce a corresponding
counterterm to the action, which we chose to be 
$A = a+c J$ \eq{A-def}. As discussed in section \ref{sec:phi3}, 
this is equivalent to an infinite shift \eq{phi-shift}
of $\phi$. Taking this into account we have
\bea
\langle \phi_{ii} \rangle 
&=& -\frac{i\la}{J^R_i}\left(-(a+c J_i)
+ (h(0) + (\und{i})\, h'(0) + \tilde h(\und i))\right) \quad  + O(\la^2),
\label{onepoint-oneloop-2}
\eea
and the condition $\langle \phi_{00}\rangle =0$ implies
\bea
a + 8 \pi^2 \theta c\, \frac{\mu^2\theta}2 &=&  h(0),  \nn\\
 c  &=& \frac{1}{8 \pi^2 \theta} \, h'(0) \, .
\eea
This is in complete agreement with \eq{c-renormaliz},
and also with \eq{a-renormaliz} taking into account
\eq{mu-renorm-pert}.
These renormalization conditions guarantee that the one-point function
$\langle \phi_{ii} \rangle$ has a well-defined 
and nontrivial limit $N\to\infty$.

\begin{figure}[htpb]
\begin{center}
\epsfxsize=2in
   \epsfbox{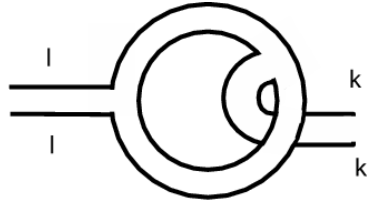}
\end{center}
 \caption{one-loop contribution to $\langle\phi_{ll}\phi_{kk}\rangle $}
\label{fig:nonplanar}
\end{figure}

\paragraph{2-point function}
Next we compute the leading contribution to the 
2-point function  $\langle\phi_{ll}\phi_{kk}\rangle$ for $l\neq k$,
which vanishes at tree level. The leading contribution 
comes from the nonplanar graph in figure \ref{fig:nonplanar},
which gives
\bea
\langle\phi_{ll}\phi_{kk}\rangle 
= \langle\phi_{kk}\rangle \langle\phi_{ll}\rangle 
+ \frac 14 \frac{(i\la)^2}{J_k\, J_l}\left(\frac{2}{J_k + J_l}\right)^2  
\label{2point-nonplanar}
\eea
(for $l\neq k$) indicating the symmetry factors, 
where the disconnected contributions 
are given by \eq{onepoint-oneloop}.
This is in complete agreement with the result 
\eq{phil-phik-kont} obtained from the 
Kontsevich model approach. Note that the counterterm $\d J$ does not
enter here.

\begin{figure}[htpb]
\begin{center}
\epsfxsize=3.5in
  \vspace{0.2in} 
   \epsfbox{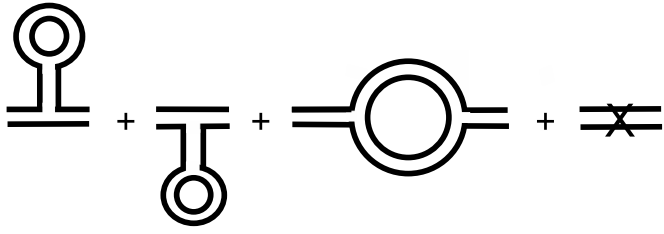}
\end{center}
 \caption{one-loop contribution to $\langle\phi_{kl}\phi_{lk}\rangle $}
\label{fig:planarprop}
\end{figure}

Similarly, the leading contribution to the 
2-point function  $\langle\phi_{kl}\phi_{lk}\rangle$ for $l\neq k$,
has the contribution indicated in figure \ref{fig:planarprop}, 
but now the counterterm $\d J$ does enter also. 
This gives the result
\bea 
\langle\phi_{kl}\phi_{lk}\rangle
&=& \frac{2}{J_k^R + J_l^R} 
- 2(i\la) \frac{\langle \phi_{kk} + \phi_{ll}\rangle}{(J_k^R + J_l^R)^2} \nn\\
&&
 + \frac{4}{(J_k^R + J_l^R)^2} \Big(\sum_{j}
  \frac{(i\la)^2}{J_k^R +J_j^R}\frac{1}{J_l^R +J_j^R}  - \frac{\d
    J_l+\d J_k}2 \Big) \,\,+ O(\la^4)
\label{2point-eom-2-pert}
\eea 
The first term is the free propagator, 
the next term the tadpole contributions, and
the last them the one-loop contribution in figure \ref{fig:planarprop}
with counterterm $\delta J$.

We  have to adjust the parameters such that the result is
well-defined and nontrivial. The last term is logarithmically
divergent,
\bea
\sum_{j} \frac{1}{J_k^R +J_j^R}\frac{1}{J_l^R +J_j^R}
&\approx& \frac 1{(8\pi^2\theta)^2}
\sum_{j}\frac{1}{(\und{j}+\mu_R^2\theta+2)^2} \nn\\
&=& -\frac 1{8\pi^2\theta} \, h'(0) 
\sim \frac 1{(8\pi^2\theta)^2} \,\log N.
\eea
Therefore the divergent terms 
$-\frac{\d J_l+\d J_k}2 - (i\la)^2 \frac 1{8\pi^2\theta} \, h'(0)$
must cancel, i.e.
\bea
-8\pi^2\theta \frac{\d\mu^2\theta}2 &=& (i\la)^2 \frac 1{8\pi^2\theta}
\, h'(0), \nn\\
\d\mu^2 &\sim&  \frac{(i\la)^2}{256\,\pi^6\theta^4}\, \log N
\label{mu-renorm-pert}
\eea
in complete agreement with \eq{mu-renormaliz}. Hence
the mass is log-divergent as expected, and no
wavefunction-renormalization $Z$ is required.

We note in particular that these one-loop computations already give
the exact results for the renormalization, as found using the 
Kontsevich model. This reflects the ``super-renormalizability'' of the
model, which is thus established rigorously.

\section{Discussion and conclusion}

We have shown that the selfdual NC
$\phi^3$ model in 4 dimensions can be renormalized and essentially solved
in terms of a genus expansion, by using the Kontsevich model.
This provides closed expressions for the free energy and certain
$n$-point functions for each genus, 
which are finite after renormalization and
valid for finite nonzero coupling. 
Remarkably, the genus 0 contribution turns out to be exact
at one loop in a special point $b = c'=0$ of moduli space.
An instability is found if the coupling constant reaches a critical
coupling, as expected for the $\phi^3$ model.

It is very remarkable that a 
nontrivial 4-dimensional NC $\phi^3$ field theory 
allows such a detailed analytical description. 
There is no commutative analog to our knowledge, 
which shows that the noncommutative world in some cases is 
more accessible to analytical methods than the commutative case.
Furthermore, these  NC $\phi^3$ field 
theories in different (even) dimensions can all be 
mapped to the same Kontsevich model $Z(M)$, for different  
$M$  with different eigenvalues and  degeneracies.
While the techniques used in this paper are
more--or--less restricted to the $\phi^3$ interaction, it is worth 
pointing out that the renormalization 
is determined by the genus 0 contribution only, which is
accessible in a wider class of models; see also 
\cite{Steinacker:2005wj,Steinacker:2005tf} in this context.

Perhaps the main gap in our treatment is the lack of control
over the {\em sum } over all genera $g$. While the contributions
for each genus are manifestly analytic in the coupling constant
$\la$, we have not shown that the sum over $g$ 
converges in a suitable sense. This would amount 
to a full construction of the model.
However, it seems very plausible that 
this is the case, and the sum defines an analytic function 
in $\la$ near the origin. 
It should be possible to establish this using 
the relation with the KdV hierarchy or the relation with topological 
gravity, which is beyond the scope of this paper. 
Another interesting extension would be the case $\Omega \neq 1$, which could
be considered as a perturbation
around $\Omega=1$, rather than around $\Omega =0$. We hope to 
study this problem in the near future.

\paragraph{Acknowledgements}

We are grateful for inspiring discussions with Edwin Langmann.
This work was supported by the FWF project P16779-N02.

\bibliographystyle{diss}

\bibliography{mainbib}

\end{document}